\documentclass[Numbered, nature]{sn-jnl}

\usepackage{graphicx}
\usepackage{dcolumn}
\usepackage{bm}
\usepackage[T1]{fontenc}
\usepackage[utf8]{inputenc}
\usepackage{amsmath}
\usepackage{mathtools}
\usepackage{amssymb}
\usepackage{color}
\usepackage{physics}
\usepackage{hyperref}
\usepackage{amsfonts}
\usepackage{geometry}

\numberwithin{equation}{section}

\begin{document}

\title[A Multi-Scale Machine Learning Framework for Coupled Chemical, Spin, and Structural Disorder in Alloys]{A Multi-Scale Machine Learning Framework for Coupled Chemical, Spin, and Structural Disorder in Alloys}

\author[1]{\fnm{Zhenyao} \sur{Fang}}\email{z.fang@northeastern.edu}

\author*[1]{\fnm{Qimin} \sur{Yan}}\email{q.yan@northeastern.edu}

\affil[1]{\orgdiv{Department of Physics}, \orgname{Northeastern University}, \city{Boston}, \postcode{02115}, \state{Massachusetts}, \country{United States}}


\abstract{
Understanding the thermodynamic properties of disordered magnetic alloys requires a unified treatment of configurational (chemical, spin, etc.) and structural degrees of freedom, which has remained beyond the scope of existing computational frameworks. Here we present a general framework that integrates machine learning models (such as graph neural networks and machine learning interatomic potentials) and statistical sampling methods (such as Monte Carlo and molecular dynamics simulations) to study the coupled chemical, spin, and structural disorder in alloys. We demonstrate the framework on body-centered-cubic Fe-Co alloys with interstitial carbon dopants, where the Fe-Co host exhibits intrinsic chemical and spin disorder, and the interstitial carbon introduces additional structural disorder through local lattice distortions, making the system a prototypical multi-disorder magnetic alloy. The framework predicts the order-to-disorder phase transition temperature and the melting temperature of Fe-Co alloy to be 1,000~K and 1,690~K, in excellent agreement with the experimentally measured values of 1,006~K and approximately 1,700~K, respectively. It also predicts the tetragonal-to-nearly-cubic structural transitions in Fe-Co-C alloy as temperature increases. These results establish the framework as a reliable tool for studying multi-disorder alloys, with applications to complex disordered systems such as high-entropy alloys, multiferroics, and spintronic devices.
}


\maketitle

\section*{Introduction\label{sec:intro}}
Disorder in crystalline materials can significantly affect their physical properties, leading to phenomena such as magnetic phase transitions~\cite{Bean62DisorderMagnetic, Armstrong25DisorderMagnetic}, enhanced mechanical strength~\cite{Pei23DisorderMechanical, Zhu23DisorderMechanical, Dsari23DisorderMechanical}, and anomalous charge transport~\cite{Shallcross17Transport, Chen21DisorderTransport, Fang25DisorderTransport}. In alloys, disorder can manifest in two coupled forms: configurational disorder, where atomic sites are occupied by different chemical species or spin orientations~\cite{Cordell21ConfigurationalDisorder, Fang24ConfigurationalDisorderGNN}, and structural disorder, arising from the local lattice distortions induced by chemical or spin inhomogeneity~\cite{Cliffe10StructrualDisorder}. While each type of disorder can be studied individually, for example, with configurational disorder sampled through Monte Carlo (MC) simulations~\cite{Yang20ConfigurationalDisorderMC, Fang24ConfigurationalDisorderGNN} and structural disorder captured through molecular dynamics (MD) simulations~\cite{Massobrio15StructuralDisorderMD}, the two are intrinsically entangled. Any change in site occupancy generates forces on neighboring atoms, coupling the configurational and structural degrees of freedom. Accurately capturing this coupling in a single computational framework remains a central challenge in the theoretical modeling of disordered materials.

Magnetic alloys represent a well-established class of materials with strong coupling of configurational (chemical and spin) and structural disorder, making them a natural system for studying the above entanglement~\cite{Kakehashi12DisorderedMagnets, Kaneyoshi98DisorderedMagnets}. Among them, the Fe-Co binary alloy is a prototypical system~\cite{Jesus16FeCo_Review, Sundar05FeCo_Review, Burkert04FeCo_Review,Sourmail05FeCo_Review, Koutsopoulos17FeCo_Review}, known for its high Curie temperature, strong ferromagnetism~\cite{Torchio13FeCo_transition}, ultra-low magnetic damping~\cite{Schoen16FeCo_damping}, and an order-to-disorder phase transition near 1,006~K~\cite{Fultz91FeCo_transition, Seehra76FeCo_transition, Oyedele77FeCo_transition}. Recent efforts have focused on adding light elements such as B~\cite{Wells89FeCoB, Charles88FeCoB} and C~\cite{Marciniak23FeCoC, Holodelshikov11FeCoC} to further tune its magnetic and electronic properties. Due to their small atomic radii and distinct electronic character relative to Fe and Co, those light elements preferentially occupy interstitial sites in the lattice, inducing significant local lattice distortions~\cite{Delczeg14FeCoC_interstitial, KhanImran15FeCoC_interstitial}. Those structural distortions, along with the distribution of carbon atoms among interstitial sites, couple with the intrinsic chemical and spin disorder of the Fe-Co host sublattice, making Fe-Co doped with light interstitial elements a challenging multi-disorder system with broad relevance to permanent magnet and spintronic applications.

Modeling such multi-disorder systems from first-principles methods is computationally challenging. Density functional theory (DFT), while accurate, is expensive for the large supercells and extensive configurational sampling required to study disorder effects. Therefore, surrogate models such as tight-binding Hamiltonians~\cite{Ashhab17TB_Model}, classical spin models (e.g., Ising and Heisenberg Hamiltonians)~\cite{Chowdhury86IsingModel}, and cluster expansion methods~\cite{Drautz19ClusterExpansion, Sanchez93ClusterExpansion} have been applied to predict the total energies for MC simulations, but these approaches are typically restricted to fixed lattice geometries and cannot capture structural distortions. Classical force fields can in principle describe the atomic dynamics through MD simulations, but few have been parameterized to simultaneously capture chemical and spin disorder in magnetic alloys. Recent developments in machine learning models such as graph neural networks (GNNs)~\cite{Chen19GNN, Fung21GNN, Reiser22GNN} and machine learning interatomic potentials (MLIPs)~\cite{Wang24MLIP, Jacobs25MLIP} provide another route to address these limitations, enabling accurate predictions of energies, forces, stresses, and magnetic moments at a fraction of the computational cost of DFT. This makes it practical to sample the vast configurational space and compute thermodynamic properties such as order-to-disorder phase transition temperature and ensemble-averaged physical properties~\cite{Fang24ConfigurationalDisorderGNN}. However, a unified framework that integrates these tools to simultaneously model the coupled chemical, spin, and structural disorder has not been established.

In this work, we present a general computational framework that integrates machine learning models (GNNs and MLIPs) into statistical sampling methods (MC and MD simulations) to model disordered alloys. In this framework, proposed MC moves (elemental swaps and spin flips) are followed by MLIP-driven MD simulations to capture structural distortions, with the GNN evaluating the configurational energy and physical properties of interest at each step. Statistical averaging over those configurations then yields thermodynamic properties such as heat capacity and ensemble-averaged properties such as total magnetic moment. 

We demonstrate this framework on body-centered-cubic Fe-Co alloys with interstitial carbon dopants. We trained a Transformer-based GNN and fine-tuned the CHGNet MLIP on a dataset of Fe-Co-C alloys with structural distortions, achieving accuracy comparable to state-of-the-art models. Without structural distortions, the GNN-driven MC simulations identify a linear carbon-chain configuration as the low-temperature ground state of Fe-Co-C, whose stability arises from the local force balance around interstitial carbon atoms. When structural distortions are incorporated through NPT-MD simulations, the framework reproduces the experimental order-to-disorder phase transition temperature of Fe-Co (predicted 1,000~K vs. experimental 1,006~K~\cite{Seehra76FeCo_transition, Oyedele77FeCo_transition}) and the experimental melting temperature (predicted 1,690~K vs. approximated 1,700~K~\cite{Muralles22FeCo_MeltingPoint, Woodcock07FeCo_MeltingPoint, Rodrigues15FeCo_MeltingPoint}). It predicts a dramatic reduction of the carbon order-to-disorder transition temperature relative to the structural-distortion-free case, and captures a tetragonal-to-nearly-cubic structural transition as carbon disorders with increasing temperature. These results demonstrate the necessity of including both configurational disorder and structural distortions when modeling complex alloys, and establish this framework as a reliable and transferable method for studying multi-disorder alloys where configurational and structural degrees of freedom are strongly coupled.

\section*{Results and Discussions}
\subsection*{Machine Learning Framework for Coupled Configurational and Structural Disorder}

\begin{figure*}[htb]
\centering
\includegraphics[width=\linewidth]{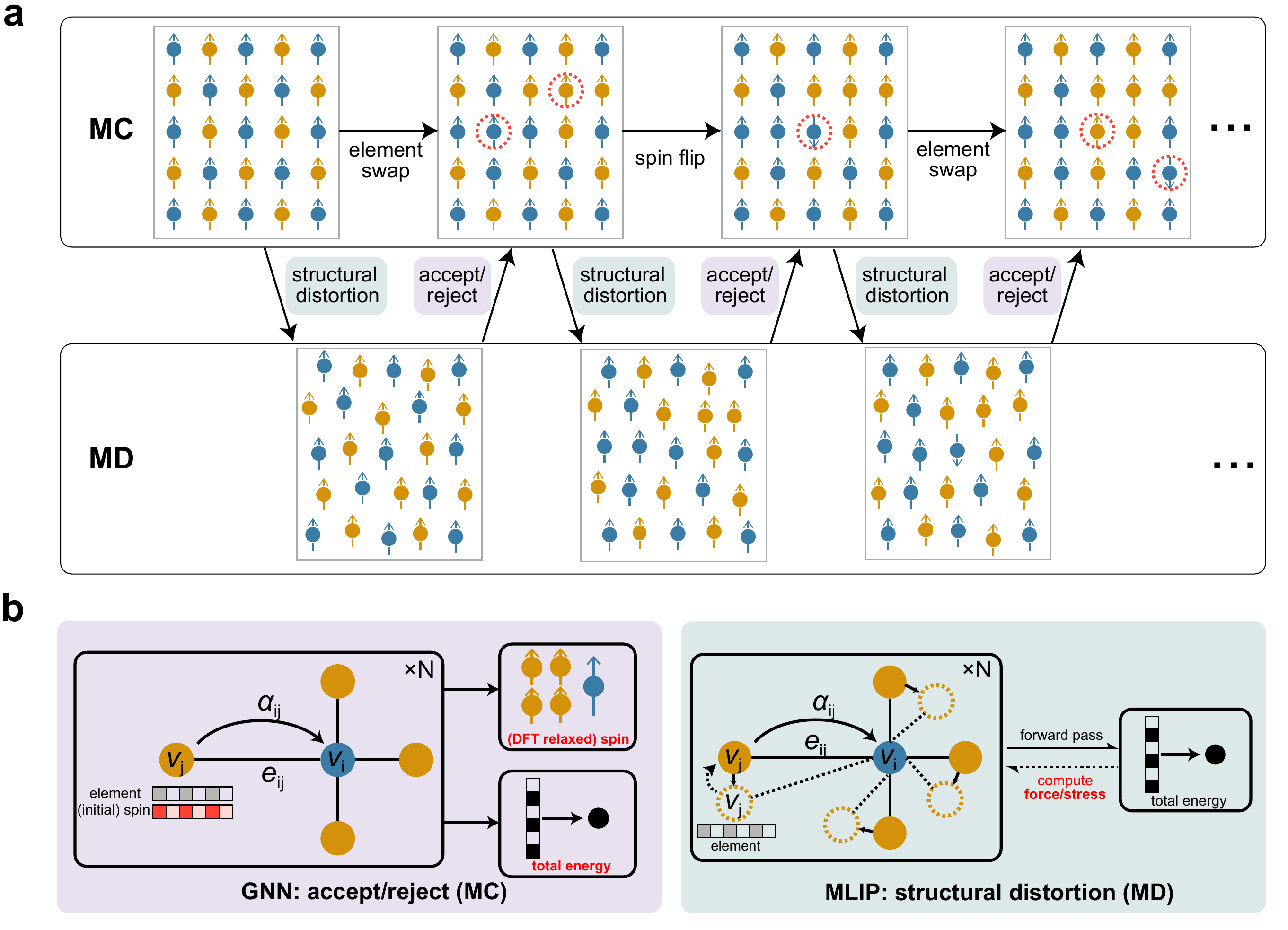}
\caption{Schematic illustration of the computational framework for modeling disordered alloys. (a) The MC+MD sampling scheme, where proposed MC moves are followed by MD simulations to capture structural distortions, and the resulting average energy is used to accept or reject the move. (b) The GNN (left) takes atomic species and initial spin configurations as node features, and predicts the site-resolved (DFT-relaxed) magnetic moments and total energy. The MLIP (right) predicts forces and stresses through backpropagation of the energy, which are used to drive the structural relaxations or MD simulations.}
\label{fig:main_workflow}
\end{figure*}

The proposed framework couples MC and MD simulations to jointly sample configurational and structural degrees of freedom, as shown in Fig.~\ref{fig:main_workflow}(a). In this framework, proposed MC moves such as elemental swaps and spin flips are applied to the current configuration, after which MD simulations are performed to capture the resulting structural distortions. A few snapshots are drawn from the MD trajectory, and their energies are averaged as the effective configurational energy, which is then used to accept or reject the proposed move through Metropolis or Wang-Landau sampling~\cite{Prokhorenko18MetropolisSampling, Wang01WLSampling, Zhou06WLSampling}. In this way, the framework samples neither discrete points in the phase space like a standard MC simulation, nor a single trajectory like a standard MD simulation, but an ensemble of trajectories that collectively capture both configurational and structural disorder.

Evaluating the configurational energy and physical properties within this framework requires efficient surrogate models for total energies, forces, stresses, and physical quantities such as magnetization. While DFT can in principle provide all these quantities, it is computationally too expensive for the numerous evaluations required during MC sampling. In this framework, we use GNNs to evaluate total energies and physical properties, and MLIPs to compute forces and stresses for MD simulations, as shown in Fig.~\ref{fig:main_workflow}(b). In a GNN, the crystal structure is represented as a graph where atoms are nodes and bonds are edges~\cite{Reiser22GNN, Fagn26GNN_Review}. Node features encode site properties such as atomic species and spin, while edge features encode bonding properties such as bond length and direction. The graph passes through several message-passing convolution layers, after which node features are decoded by a multilayer perceptron to predict node-level properties such as site-resolved magnetic moments, and pooled into a graph-level feature vector to predict global properties such as total energy. More details on the specific GNN architecture employed in this work are provided in the following.

For magnetic alloys, the GNN should account for the fact that DFT produces two sets of magnetic moments: the initial guess $\{ s_{i, \text{initial}} \}$ supplied as input, and the self-consistently relaxed moments $\{ s_{i, \text{relaxed}} \}$ that reflect interatomic magnetic interactions under a given initial guess, where $i$ is the atomic site index. It is important to note that $\{ s_{i, \text{relaxed}} \}$ depends on the initial guess and is typically not the same as the ground-state magnetic configuration, and they (rather than the initial guess) are required for computing spin-related properties such as total magnetic moment. We therefore include both atomic species (one-hot encoded) and initial spin $\{ s_{i, \text{initial}} \}$ (through an embedding layer) as node features, and train the GNN to simultaneously predict $\{ s_{i, \text{relaxed}} \}$ as node-level outputs and total energy as a graph-level output. The importance of including those initial spin features in the GNN model is demonstrated through ablation studies in Supplementary Information 2.2.

Although forces and stresses can also be predicted directly by the GNN (as node-level vector property and as graph-level tensorial property, respectively), they must be calculated as derivatives of total energy with respect to atomic positions and cell parameters in order to ensure a conservative force field. In this work, we adopt CHGNet~\cite{Deng23chgnet} as the MLIP for structural dynamics, which computes forces and stresses through backpropagation of the total energy and was developed as a universal potential for materials and performs well on our dataset after fine-tuning. We note that while CHGNet predicts energies and magnetic moments, it does so by relaxing to the lowest-energy magnetic configuration for a given atomic structure, effectively predicting the ground-state magnetic energy and moments instead of the energy and moments $\{ s_{i, \text{relaxed}} \}$ that correspond to a specific initial configuration $\{ s_{i, \text{initial}} \}$. Since our simulations require evaluations for arbitrary spin configurations, we therefore use CHGNet exclusively for force and stress evaluation, and use the GNN for total energy and magnetic moment predictions. Furthermore, benchmark DFT calculations demonstrate that magnetic configurations do not significantly affect interatomic forces and stresses in our system (see Supplementary Information 1.3), justifying the use of a non-magnetic MLIP for structural dynamics in this work.

\subsection*{Application to Fe-Co-C Alloys}

We demonstrate the framework on body-centered-cubic Fe-Co alloys doped with interstitial carbon atoms. As shown in Supplementary Information 1.1, DFT calculations confirm that carbon preferentially occupies interstitial sites with a formation energy of around 2.3~eV, significantly lower than the substitutional formation energy of 3.98~eV and 4.60~eV for replacing an Fe or Co atom. Such interstitial dopants introduce significant lattice distortions and local atomic displacements, making Fe-Co-C an ideal model system for studying the coupling of chemical, spin, and structural disorder and demonstrating the capability of the framework. We note that while this work focuses on interstitial carbon dopants, the framework can be readily extended to other cases, such as substitutional dopants, vacancies, and anti-site defects, making it broadly applicable to a wide range of doped materials or high-entropy alloys.

\subsubsection*{Training and Validation of GNN and MLIP Models}

\begin{figure*}[htb]
\centering
\includegraphics[width=\linewidth]{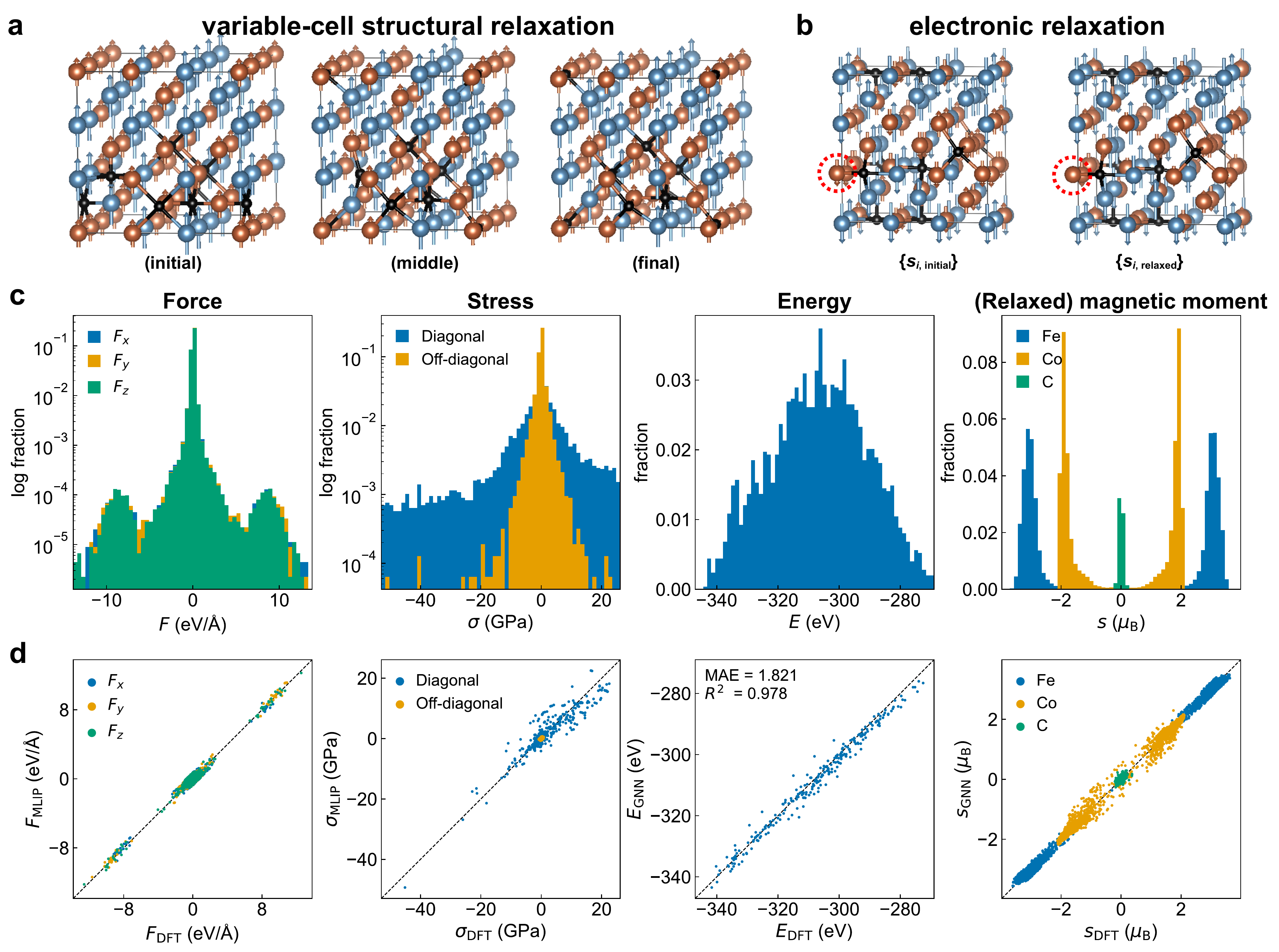}
\caption{Dataset construction and model training results. (a) An example variable-cell structural relaxation trajectory  from the MLIP fine-tuning dataset, illustrated by three snapshots (initial, middle, and final). All snapshots have the FM magnetic configuration. Color code: Fe (blue), Co (brown), and C (black). (b) An example configuration from the GNN training dataset, showing the initial magnetic moments $\{ s_{i, \text{initial}} \}$ and the self-consistently relaxed magnetic moments $\{ s_{i, \text{relaxed}} \}$ after DFT electronic relaxation. The red dashed circle highlights a site where the relaxed magnetic moment differs significantly from the initial guess. (c) Distributions of the training targets: force components $F_x$, $F_y$, $F_z$ (eV/\AA) and diagonal/off-diagonal stress components (GPa) from the MLIP dataset, and total energy (eV) and relaxed site-resolved magnetic moments ($\mu_B$) for the GNN dataset. Force and stress distributions are shown on a logarithmic scale. (d) Parity plots comparing DFT reference values against model predictions for forces and stresses (MLIP), and total energies and site-resolved magnetic moments (GNN).}
\label{fig:main_GNN_MLIP_training}
\end{figure*}

We constructed two datasets for training the MLIP and GNN models, respectively, both based on $3 \times 3 \times 3$ supercell Fe-Co alloys doped with various amounts of interstitial carbon atoms. The ratio of Fe to Co is randomly chosen between 0 and 1, covering the full range of Fe-Co compositions, and the carbon concentration $x_C$ ranges from 0\% to 11.1\%, corresponding to zero to six carbon atoms per supercell.

For the MLIP dataset, we selected 500 atomic configurations with ferromagnetic (FM) magnetic ordering and performed variable-cell DFT relaxation on each configuration. The dataset encompasses a wide range of structural environments, including configurations where interstitial carbon atoms hop between neighboring interstitial sites during relaxation, as illustrated by the example trajectory in Fig.~\ref{fig:main_GNN_MLIP_training}(a). To ensure that the extracted snapshots are representative of the full relaxation trajectory and cover a wide range of forces and stresses, we applied farthest point sampling~\cite{Trestman26FPS} to extract 5 snapshots per trajectory, yielding 2,500 snapshots in total. The distributions of force and stress components are shown in Fig.~\ref{fig:main_GNN_MLIP_training}(c). Forces are concentrated near 0~eV/\AA~with a small fraction of large forces near $\pm 10$~eV/\AA~arising from the early stages of relaxation, while diagonal stress components exhibit a broader distribution than off-diagonal components.

We fine-tuned the pre-trained CHGNet model on this dataset; further fine-tuning details are provided in the Methods section and Supplementary Information 2.1. Before fine-tuning, the pre-trained model accurately predicts forces and off-diagonal stress components, but cannot reproduce the diagonal stress components (see Supplementary Figure 5). After fine-tuning, the mean absolute error (MAE) for forces and stresses is 0.119~eV/\AA~and 0.443~GPa, with $R^2$ coefficients of 0.887 and 0.916, respectively, as shown in Fig.~\ref{fig:main_GNN_MLIP_training}(d). Specifically, the diagonal components are significantly improved after fine-tuning. The performance of the fine-tuned model is comparable to state-of-the-art MLIP models~\cite{Deng23chgnet, Yang25MLIP_performance}, demonstrating its reliability for molecular dynamics simulations.

For the GNN dataset, we randomly selected 500 snapshots from the MLIP dataset, and generated 8 configurations per snapshot by fixing the lattice and atomic positions, while randomly swapping Fe-Co atomic pairs and assigning spin-up and spin-down orientations to each atomic site, yielding 4,000 configurations covering a wide range of atomic environments and magnetic configurations. For each configuration, we performed DFT self-consistent calculations to obtain the relaxed magnetic moments $\{ s_{i, \text{relaxed}} \}$. As illustrated by the example in Fig.~\ref{fig:main_GNN_MLIP_training}(b), $\{ s_{i, \text{relaxed}} \}$ can differ significantly from the initial guess $\{ s_{i, \text{initial}} \}$, for example at sites highlighted by the red dashed circle. The initial magnetic moments were set to $\pm 3~\mu_B$ for Fe, $\pm 2~\mu_B$ for Co, and $0~\mu_B$ for C atoms. The distribution of total energy and relaxed magnetic moments is shown in Fig.~\ref{fig:main_GNN_MLIP_training}(c). After electronic relaxation, the magnetic moment distributions for each species broaden but remain concentrated near their initial values.

We trained a GNN based on the Transformer convolution layers~\cite{Shi21Transformer} to predict the total energy as a graph-level output and site-resolved relaxed magnetic moments $\{ s_{i, \text{relaxed}} \}$ as node-level outputs; further training details can be found in the Methods section and Supplementary Information 2.2. The trained model achieves MAEs of 1.821~eV/cell (34.0~meV/atom) and 0.065~$\mu_B$ for energy and magnetic moments, with $R^2$ coefficients of 0.978 and 0.999, respectively, as shown in Fig.~\ref{fig:main_GNN_MLIP_training}(d). The energy MAE per atom is comparable to previous GNN models for predicting magnetic material properties~\cite{Xu25GNN_performance}, demonstrating the reliability of the model for thermodynamic sampling.

Ablation studies are provided in Supplementary Information 2.2. Omitting the initial spin $\{ s_{i, \text{initial}} \}$ from the node features leads to model collapse, where the GNN predicts only the mean magnetic moments and fails to reproduce the correct site-resolved distribution; this confirms the necessity of including initial spin information in our GNN model. Additional experiments with persistent homology features~\cite{Jiang21persistent_homology, Fang25Persistent} and an equivariant neural network architecture~\cite{Batzner22EquivariantGNN} (see Supplementary Information 2.2) showed no significant improvement over the Transformer-based GNN in this system, and were therefore not adopted. 

\subsubsection*{Phase Transitions Without Structural Distortions}
With the trained models, we first study the phase transitions in Fe-Co-C alloys without structural distortions, where atomic positions and lattice parameters are held fixed and the framework reduces to a GNN-driven MC simulation sampling only configurational degrees of freedom. Details about MC setup can be found in the Methods section. 

\begin{figure}[htb]
\centering
\includegraphics[width=\linewidth]{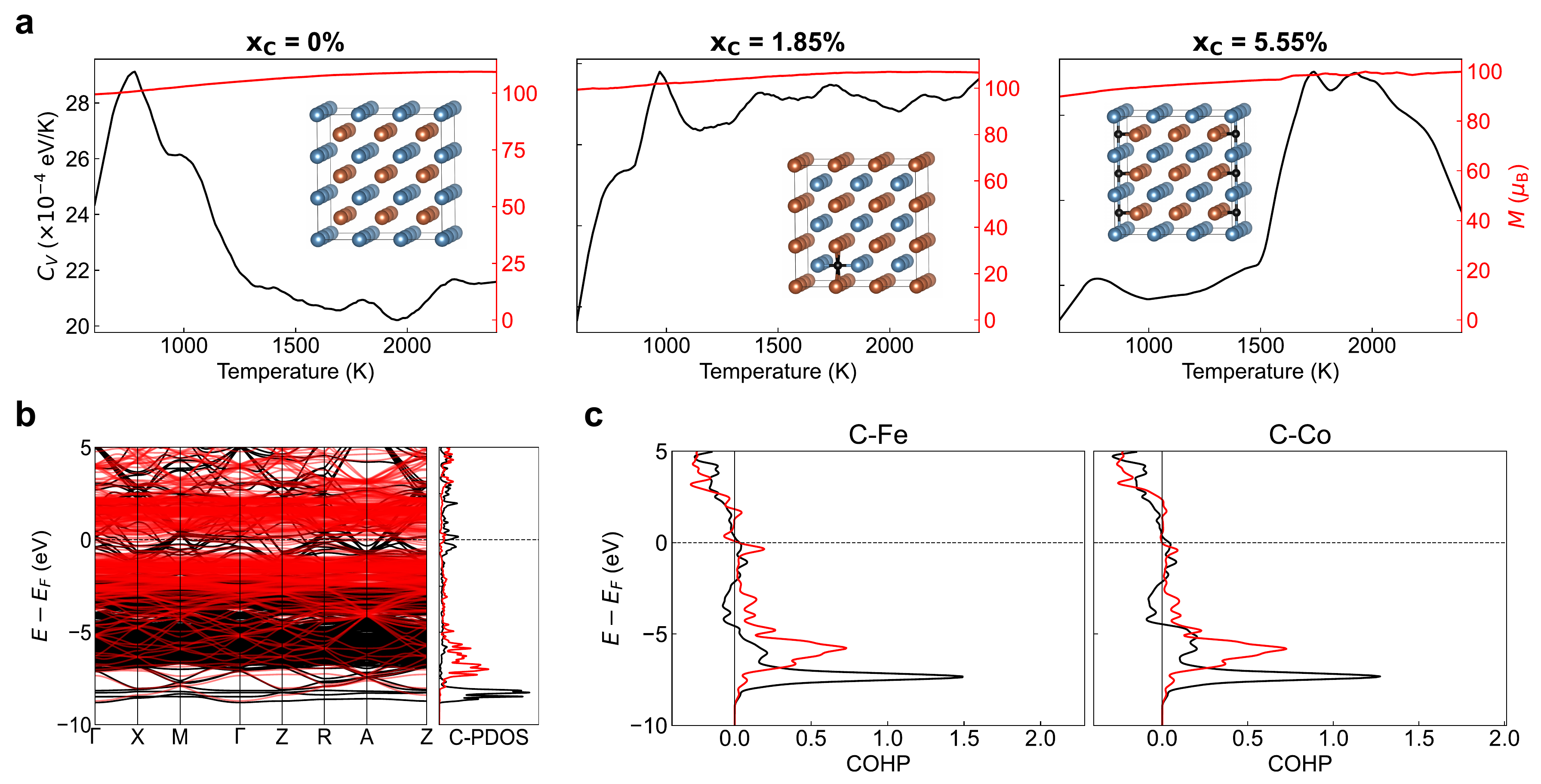}
\caption{MC simulation results for Fe-Co-C without structural distortions. (a) Heat capacity (left axis, black) and ensemble-averaged total magnetic moment (right axis, red) as a function of temperature for $x_C=$ 0\%, 1.85\%, and 5.55\%, corresponding to 0, 1, and 3 carbon atoms per $3 \times 3\times 3$ supercell, respectively. The inset shows the lowest-energy configuration identified during the MC simulation. Color code: Fe (blue), Co (brown), and C (black). (b) Band structure and carbon-projected density of states (C-PDOS) of the $C_{i1}$ interstitial configuration, with spin-up and spin-down channels shown in black and red, respectively. The Fermi level is set to zero. (c) COHP analysis for the nearest-neighboring C-Fe and C-Co atomic pairs within the $C_{i1}$ interstitial configuration. Positive and negative COHP values indicate bonding and antibonding states, respectively.}
\label{fig:main_GNNonly}
\end{figure}

We studied $3 \times 3 \times 3$ supercells of Fe-Co (50:50) with carbon concentrations of $x_C=$~0\%, 1.85\%, and 5.55\%, corresponding to 0, 1, and 3 carbon atoms per supercell, respectively. In the case of pure Fe-Co ($x_C = 0\%$, Fig.~\ref{fig:main_GNNonly}(a) left panel), the heat capacity exhibits a prominent peak at 780~K, indicating the order-to-disorder phase transition. Throughout the entire temperature range, the ensemble-averaged magnetization remains positive and close to the FM value of 4.9~$\mu_B$~per unit cell, confirming that the transition is driven by chemical disorder rather than spin disorder. The lowest-energy configuration identified during the MC simulation, shown in the inset of the panel, corresponds to the perfectly ordered FeCo structure.

In the case of one carbon atom per supercell ($x_C = 1.85\%$, Fig.~\ref{fig:main_GNNonly}(a) middle panel), the carbon configurational space contains only two symmetry-inequivalent interstitial configurations $C_{i1}$ and $C_{i2}$, with the same multiplicity (see Supplementary Information 1.1), effectively reducing the carbon sector of the configurational space to a two-level system. If the carbon and Fe-Co configurational degrees of freedom were decoupled, such a two-level system would give rise to a Schottky anomaly peaked at $T \approx 0.42 \Delta E / k_\text{B} \approx 292$~K~\cite{Tari03SchottkyAnomaly}, where $\Delta E \approx 0.06$~eV is the DFT energy difference between the $C_{i1}$ and $C_{i2}$ interstitial sites (see Supplementary Information 3.2). However, the simulated heat capacity instead exhibits a broad peak at significantly higher temperatures. This indicates that the carbon and Fe-Co configurational degrees of freedom are strongly coupled, where the Fe-Co chemical disorder effectively renormalizes the carbon transition temperature far above its intrinsic two-level value. 

For three carbon atoms per supercell ($x_C = 5.55\%$, Fig.~\ref{fig:main_GNNonly}(a) right panel), the heat capacity shows a prominent peak above 1,700~K, significantly higher than the pure Fe-Co case, indicating that C-C interactions significantly increase the order-to-disorder phase transition temperature. The ensemble-averaged magnetization remains positive throughout the entire temperature range, confirming that this transition is also dominated by chemical disorder. This is further confirmed by fixed-spin MC simulations which show nearly identical heat capacity (see Supplementary Information 3.1). The lowest-energy configuration identified during the MC simulation, shown in the inset of the panel, features three carbon atoms forming a linear chain. This chain configuration is energetically favorable because the forces exerted by each interstitial carbon are aligned along the chain axis and partially cancel each other, whereas non-collinear carbon configurations lead to competing forces in different directions that raise the total energy (see Supplementary Information 3.1 for an example of a high-energy configuration during the MC simulation). 

To further characterize the electronic structure of the $C_{i1}$ interstitial configuration, we calculated its band structure and carbon-projected density of states (C-PDOS), as shown in Fig.~\ref{fig:main_GNNonly}(b). The C-PDOS shows that carbon does not contribute to the electronic states near the Fermi level, indicating that the interstitial carbon does not significantly perturb the electronic structure of the Fe-Co host. This is further supported by crystal orbital Hamilton population (COHP) analysis~\cite{COHP} (Fig.~\ref{fig:main_GNNonly}(c)), which shows that the bonding and antibonding states formed between carbon and neighboring Fe or Co atoms lie significantly below and above the Fermi level, respectively. Furthermore, the C-Fe and C-Co interactions are nearly identical in strength, consistent with the small energy difference of $\Delta E \approx 0.06$~eV between the two interstitial sites $C_{i1}$ and $C_{i2}$. Band structures for all three carbon concentrations, and full COHP analysis for the interstitial configuration, are provided in Supplementary Information 1.2.

\subsubsection*{Phase transitions With Structural Distortions}

We now incorporate structural distortions into the MC framework through NPT-MD simulations, allowing the atomic positions and lattice parameters to evolve at each MC step. Results using NVT-MD simulations can be found in Supplementary Information 4.2. NPT-MD is adopted as the primary approach as it correctly accounts for the thermodynamic response of the lattice, including thermal expansion and the structural transitions accompanying changes in carbon ordering.

\begin{figure}[htb]
\centering
\includegraphics[width=\linewidth]{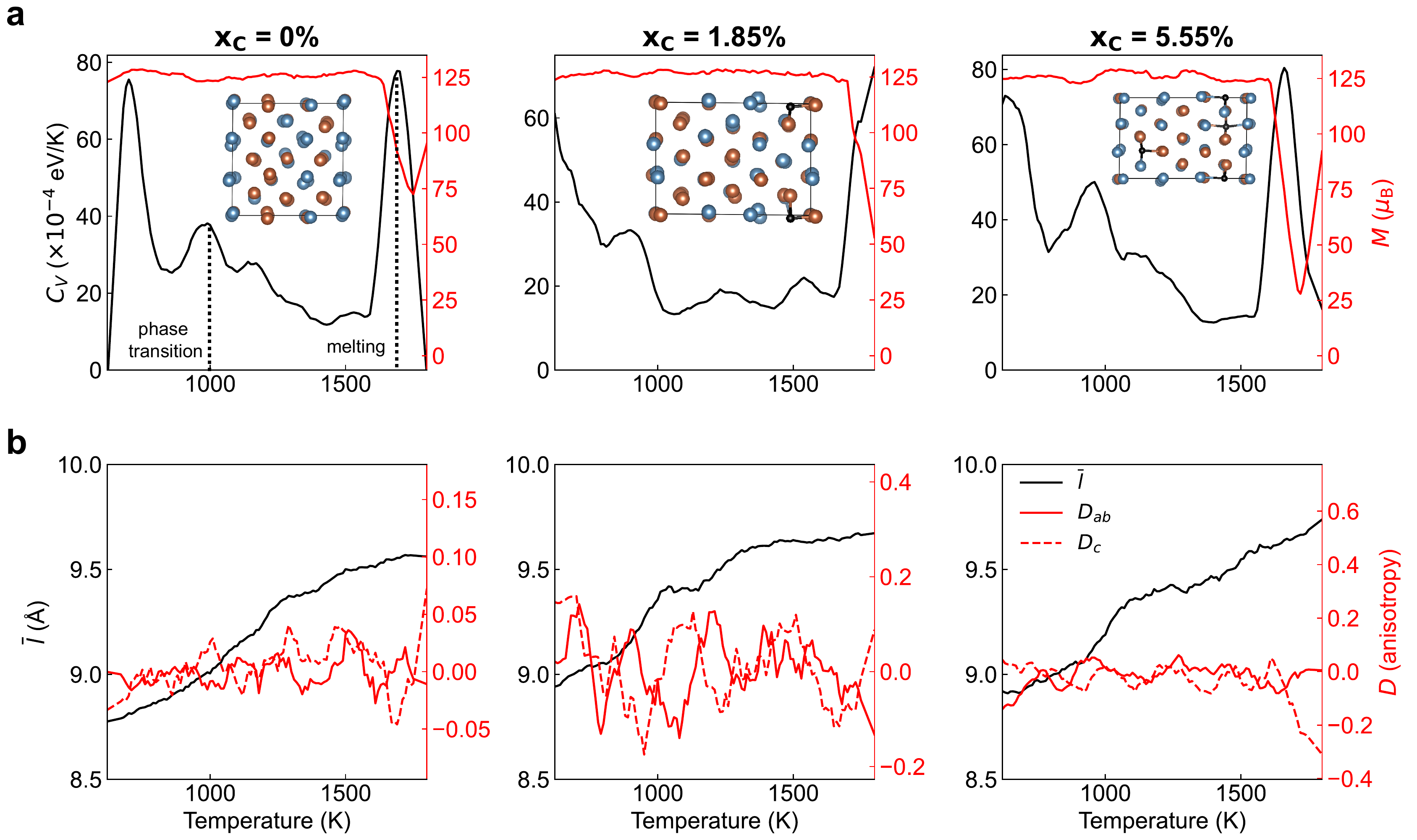}
\caption{MC simulation results for Fe-Co-C with structural distortions captured by NPT-MD. (a) Heat capacity (left axis, black) and ensemble-averaged total magnetic moment (right axis, red) as a function of temperature for $x_C=$ 0\%, 1.85\%, and 5.55\%, corresponding to 0, 1, and 3 carbon atoms per $3 \times 3\times 3$ supercell, respectively. Color code: Fe (blue), Co (brown), and C (black). (b) Isotropic lattice order parameter $\bar{l}$ (left axis, black) and anisotropic lattice order parameters $D_{ab}$ (right axis, solid red) and $D_c$ (right axis, dashed red) as a function of temperature for the same three carbon concentrations.}
\label{fig:main_MD_NPT}
\end{figure}

The heat capacity and ensemble-averaged magnetization are shown in Fig.~\ref{fig:main_MD_NPT}. In the case of pure Fe-Co ($x_C= 0\%$, Fig.~\ref{fig:main_MD_NPT}(a) left panel), the heat capacity exhibits a peak at 1,000~K, in excellent agreement with the experimentally measured order-to-disorder phase transition temperature of 1,006~K~\cite{Seehra76FeCo_transition, Oyedele77FeCo_transition}, validating that the framework accurately reproduces the experimental thermodynamic properties. 
When $x_C$ increases to 5.55\% (Fig.~\ref{fig:main_MD_NPT}(a) right panel), the lowest-energy configuration no longer contains the carbon chain, in contrast to the results without structural distortions. In the absence of structural distortions, the carbon chain is stabilized by the cancellation of forces along the chain axis. But when structural distortions are introduced, the lattice locally distorts around each carbon atom, breaking the force cancellation symmetry and the energetic advantage of the chain configuration. This interpretation is further supported by MC simulations with NVT-MD calculations, where the lowest-energy configuration similarly does not contain a carbon chain (see Supplementary Information 4.2).

Comparing the heat capacity across all three carbon concentrations, a prominent peak near 1,000~K is observed in all cases, in contrast to the case without structural distortions where the peak shifts to higher temperatures with increasing carbon concentration. This robustness of the 1,000~K peak confirms that it is associated with Fe-Co chemical ordering, instead of with carbon-ordering. Besides, once structural distortions are introduced, the carbon chain configuration is no longer energetically favorable, removing the C-C interaction contribution that raises the carbon transition above 1,000~K in the case without structural distortions. This demonstrates that structural distortions play a crucial role in correctly capturing the thermodynamic properties of disordered alloys, and that neglecting them could lead to a significant overestimation of the phase transition temperature.

At higher temperatures, all three heat capacities show a peak associated with alloy melting, as confirmed by structural analysis shown in Supplementary Information 4.3 and Supplementary Figure 13. For pure Fe-Co, the predicted melting temperature of 1,690~K is in good agreement with the experimental value of approximately 1,700~K~\cite{Muralles22FeCo_MeltingPoint, Woodcock07FeCo_MeltingPoint, Rodrigues15FeCo_MeltingPoint}, further validating the accuracy of the framework. We note that melting is observed only in NPT-MD simulations and not in NVT-MD calculations. This is physically consistent with the fact that NVT-MD constrains the cell volume, preventing the thermal expansion that accompanies melting. This also highlights the importance of NPT-MD simulations for correctly capturing high-temperature behavior in disordered alloys.

To quantify the evolution of lattice geometry with temperature, we define the isotropic lattice order parameter $\bar{l} = (a + b+c)/3$ and two anisotropic lattice order parameters inspired by the symmetry-adapted strain decomposition for cubic systems~\cite{Eckstein22SymmetricStrain, Carpenter10SymmetricStrain, Carpenter98SymmetricStrain}
\begin{align}
    &D_{ab} = \frac{a - b}{\sqrt{2}\,\bar{l}} \\
    &D_c = \frac{a + b - 2c}{\sqrt{6}\,\bar{l}}
\end{align}
where $\bar{l}$ is the normalization factor instead of a fixed reference lattice constant and accounts for the thermal expansion at finite temperature. Here $D_{ab}$ and $D_c$ form an orthogonal basis for the anisotropic subspace of lattice distortions, and any deviation from cubic symmetry that preserves orthogonal lattice angles can be expressed as a linear combination of them. Further details and explicit pairwise lattice constant differences in terms of $D_{ab}$ and $D_c$ are provided in Supplementary Information 4.3. Their temperature dependence is shown in Fig.~\ref{fig:main_MD_NPT}(b), and the full evolution of individual lattice constants and angles is provided in Supplementary Figure 12.

In all three carbon concentrations, the isotropic order parameter $\bar{l}$ increases monotonically with temperature, reflecting the lattice thermal expansion. At low temperatures, the anisotropic order parameters show small deviations from zero, indicating slight tetragonal or orthorhombic distortions of the lattice.  As temperature increases, both anisotropic order parameters fluctuate near zero, indicating a transition to a nearly-cubic structure. This tetragonal-to-nearly-cubic transition is consistent with the disordering of the carbon sublattice at elevated temperatures, which removes the symmetry-breaking effect of the carbon chain on the host lattice. 

Furthermore, for $x_C = 1.85\%$ (Fig.~\ref{fig:main_MD_NPT}(b) middle panel), the fluctuations in $D_{ab}$ and $D_c$ are significantly larger than those for $x_C = 0\%$ and $x_C = 5.55\%$. This is related to the two-level nature of the carbon configurational space at this concentration. With only two symmetry-inequivalent interstitial configurations $C_{i1}$ and $C_{i2}$ available, the system alternates between two structurally distinct states during the simulation, each inducing a different lattice distortion. This switching between two distorted states produces large fluctuations in the anisotropic order parameters, in contrast to the $x_C = 0\%$ case where no carbon-induced symmetry breaking occurs, and the $x_C = 5.55\%$ case where the larger configurational space averages out the distortions more effectively.

Together, these results demonstrate that structural disorder plays an important role in the thermodynamic properties of Fe-Co-C alloys. It destabilizes the carbon chain ground state, suppresses the carbon order-to-disorder phase transition temperature, and produces the tetragonal-to-nearly-cubic structural transition as carbon disorders with increasing temperature. More broadly, these findings demonstrate that configurational and structural disorder must be taken into consideration simultaneously to correctly predict the thermodynamic properties of disordered alloys, and this can be addressed by the general computational framework involving machine learning models and statistical sampling methods presented in this work. This framework reproduces experimental benchmarks for Fe-Co-C with first-principles accuracy and opens a path towards modeling the thermodynamics of complex disordered materials at experimentally relevant length and time scales, and can be extended to study other physical properties such as magnetic anisotropy and magnetostriction, or applied to other multi-disorder systems such as high-entropy alloys and permanent magnet materials.

\section*{Methods}
\subsection*{First-Principles Calculations}
First-principles calculations were performed using DFT as implemented in the Vienna Ab initio Simulation Package (VASP)~\cite{Kresse93VASP1, Kresse96VASP2}. We used the projector augmented wave pseudopotentials and the Perdew-Burke-Ernzerhof functional in the generalized gradient approximation for all calculations~\cite{Kresse99PAW1, Blochl94PAW2, Perdew96GGA}. A plane-wave kinetic energy cutoff of 550~eV was used, with an electronic energy convergence threshold of $10^{-6}$~eV and an ionic force convergence threshold of 0.01~eV/\AA. The Brillouin zone was sampled using a $\mathbf{k}$-point grid with a density of 0.03 $2 \pi$/\AA. We used the DFT+U method to account for the electron correlation effect~\cite{Anisimov91DFT_U, Cococcioni05DFT_U}, where Hubbard $U$ values of 4~eV and 3~eV were applied to the $3d$ orbitals of Fe and Co atoms, respectively. All calculations were performed with spin polarization.

\subsection*{Dataset Construction}
We constructed two datasets for training the MLIP and GNN models, based on $3 \times 3 \times 3$ supercells of body-centered-cubic Fe-Co alloys doped with interstitial carbon atoms. The Fe concentration $x_\text{Fe}$ was randomly chosen across the full composition range (with $x_\text{Co} = 1 - x_\text{Fe}$), and the carbon concentration $x_C$ ranges from 0\% to 11.1\%, corresponding to zero to six carbon atoms per supercell. All carbon atoms were randomly placed at interstitial sites. 

For the MLIP dataset, 500 atomic configurations were randomly selected with FM magnetic ordering, and variable-cell DFT relaxation was performed on each configuration. To ensure that the extracted snapshots are representative of the full relaxation trajectory and cover a wide range of forces and stresses, farthest point sampling was applied to extract 5 snapshots per trajectory~\cite{Trestman26FPS}. Specifically, the sampling was performed in the space of normalized force vectors, iteratively selecting snapshots that are maximally distant from all previously chosen snapshots in this feature space. This maximizes the diversity of force and stress distributions in the dataset. The first and last snapshots of each trajectory were always included to ensure coverage of both the highly distorted initial configuration and the relaxed final structure. In total, this dataset contains 2,500 snapshots, and was split into training, validation, and test sets with a ratio of 80/10/10.

For the GNN dataset, 500 snapshots were randomly selected from the MLIP dataset. For each snapshot, the lattice and atomic positions were fixed, and 8 configurations were generated by randomly swapping Fe-Co atomic pairs and assigning spin-up and spin-down orientations to each atomic site, with initial magnetic moments set to be $\pm 3~\mu_B$ for Fe, $\pm 2~\mu_B$ for Co, and $0~\mu_B$ for C atoms. Self-consistent calculations were then performed on each configuration to obtain the total energy and the relaxed magnetic moments. In total, this dataset contains 4,000 configurations, and was split into training, validation, and test sets with a ratio of 80/10/10.

\subsection*{Training of GNN and MLIP Models}
Our GNN model adopts a multi-task architecture that simultaneously predicts the total energy as a graph-level output and site-resolved magnetic moments as node-level outputs~\cite{Chen19GNN, Fung21GNN}. The input crystal structure is converted into a graph, where atoms and chemical bonds are represented as nodes and edges, respectively. The node features are constructed from one-hot encodings of the atomic species and an embedding of the initial magnetic moments, while the edge features are constructed from interatomic distances expanded in a Gaussian basis. The graph passes through several Transformer convolution layers where neighboring nodes exchange messages through attention-based message-passing functions~\cite{Shi21Transformer}. After the convolution layers, the node features are decoded by a multilayer perceptron to predict the site-resolved magnetic moments, and also pooled into a graph-level feature vector which is then decoded by a separate multilayer perceptron to predict the total energy. 

The model was trained by minimizing a combined loss function consisting of the MAE for the total energy and the MAE for the site-resolved magnetic moments, weighted equally. The model with the minimum validation loss was selected for evaluation on the test set. Hyperparameters including the number of convolution layers, the number of embedding channels, the learning rate, the weight decay, and the batch size were optimized using Bayesian optimization as implemented in Optuna~\cite{Optuna}.

For the MLIP, the pre-trained CHGNet model was fine-tuned on the MLIP dataset to predict interatomic forces and stresses. Since CHGNet is pre-trained on a large dataset of DFT calculations and already provides reasonable force predictions for a wide range of materials, this fine-tuning focuses on improving its performance for the specific chemical environments in Fe-Co-C alloys. The model was fine-tuned by minimizing a combined loss function of mean squared error (MSE) for forces and stresses, with equal weighting between the two targets. Similar to GNN model training, we performed hyperparameter optimization using Bayesian optimization as implemented in Optuna~\cite{Optuna}, where the hyperparameters include the learning rate, batch size, and the set of trainable modules. The latter controls which layers of the pre-trained model are unfrozen during fine-tuning, ranging from only the site-wise output layers to progressively deeper atom convolution, bond convolution, and angle layers.
 
\subsection*{MC and MD Simulations}
MC simulations were performed using the Metropolis sampling algorithm~\cite{Prokhorenko18MetropolisSampling}, where proposed moves consist of elemental swaps between neighboring Fe and Co sites, hopping of interstitial carbons, and spin flips at randomly selected atomic sites (for Fe and Co only). At each temperature, $10^5$ MC steps were performed, with the first 25\% of steps discarded and the later 75\% steps used to compute the thermodynamic properties. The heat capacity was calculated as 
\begin{align}
    C_v = \frac{\langle E^2 \rangle - \langle E \rangle^2}{k_\text{B} T^2} \label{eqn:heat_capacity}
\end{align}
where $\langle E \rangle$ and $\langle E^2 \rangle$ are the average energy and average squared energy, respectively.

For simulations incorporating structural distortions, each MC step consists of a short MD trajectory from which 5 snapshots were drawn and their GNN-evaluated energies were averaged as the effective configurational energy; the total number of MC steps per temperature is $2 \times 10^4$. NVT-MD simulations were performed using a Berendsen thermostat~\cite{Berendsen84MD} with a timestep of 2~fs, 250 burn-in steps, and 500 production steps (thus 1~ps of MD trajectory) per MC step. NPT-MD simulations were performed using a Berendsen thermostat and a Berendsen barostat~\cite{Berendsen84MD} at a pressure of $1.01 \times 10^{-4}$~GPa, with a pressure time constant of $\tau_P=20$~fs, a temperature time constant of $\tau_T = 10$~fs, a compressibility of $6.25\times 10^{-3}$~GPa$^{-1}$. Similar to NVT-MD, we chose a timestep of 2.0~fs, 250 burn-in steps, and 500 production steps per MC step.



\section*{Acknowledgments}
Z.F. and Q.Y. thank helpful discussions with Dr. Mingzhong Wu. This work is supported by the U.S. Department of Energy, Office of Science, Basic Energy Sciences, under Award No. DE-SC0023664. This research used resources of the National Energy Research Scientific Computing Center (NERSC), a U.S. Department of Energy Office of Science User Facility located at Lawrence Berkeley National Laboratory, operated under Contract No. DE-AC02-05CH11231 using NERSC award BES-ERCAP0029544.

\section*{Author contributions statement}

Z.F. and Q.Y. conceived the experiments. Z.F. wrote the code for model training and simulations Q.Y. supervised the study. The manuscript was written through contributions from all authors, and all authors have given approval to the final version of the manuscript.

\section*{Competing interests}
The authors declare no competing interests.

\bibliographystyle{naturemag}
\bibliography{bibliography}

\end{document}